# Influence of Pre- and Post-compensation of Chromatic Dispersion on Equalization Enhanced Phase Noise in Coherent Multilevel Systems

Gunnar Jacobsen[1,*], Marisol Lidón[2], Tianhua Xu[1,2,3], Sergei Popov[2], Ari T. Friberg[2] and Yimo Zhang[3]

[1]Acreo AB, Electrum 236, SE-16440, Kista, Sweden

[2]Royal Institute of Technology, Stockholm, SE-16440, Sweden

[3]Tianjin University, Tianjin, 300072, P.R. China

**Abstract**: In this paper we present a comparative study in order to specify the influence of equalization enhanced phase noise (EEPN) for pre- and post-compensation of chromatic dispersion in high capacity and high constellation systems. This is – to our knowledge – the first detailed study in this area for pre-compensation systems. Our main results show that the local oscillator phase noise determines the EEPN influence in post-compensation implementations whereas the transmitter laser determines the EEPN in pre-compensation implementations. As a result of significance for the implementation of practical longer-range systems it is to be emphasized that the use of chromatic dispersion equalization in the optical domain – e.g. by the use of dispersion compensation fibers – eliminates the EEPN entirely. Thus, this seems a good option for such systems operating at high constellations in the future.

**Keywords.** Equalization enhanced phase noise (EEPN), chromatic dispersion compensation, quadrature phase shift keying (QPSK), pre-compensation, post-compensation

## 1. Introduction

Fiber impairments, such as chromatic dispersion (CD) and phase noise, severely impact the performance of high speed optical fiber transmission systems [1,2]. Digital coherent receivers allow complete equalization of chromatic dispersion (a linear transmission impairment) in the electrical domain by using discrete signal processing (DSP) techniques, and have become a promising alternative approach to the use of dispersion compensation fibers [3]. For systems using post-compensation of dispersion (in the receiver (Rx)) there is a complicated interplay between the discrete chromatic dispersion compensation and the laser phase noise. When it comes to the system impact of phase noise [4-13], it is important to understand that moving to higher constellations (which is a way of increasing the system capacity using the same symbol rate) tends to increase the phase noise impact. This means that higher constellation systems (i.e. realized as n-level PSK or QAM implementations with n=4,8,16,64,…) put more stringent requirements to the spectral purity of the system transmitter (Tx) and local oscillator (LO) lasers. This can be further enhanced by the total EEPN effect for systems using discrete signal processing for chromatic dispersion compensation. Due to the existence of EEPN, the requirement of laser linewidth cannot be generally relaxed for the transmission system with higher symbol rate. This interplay leads to a combination of correlated equalization enhanced phase noise (EEPN), amplitude noise and time jitter [4-13]. The detailed effect is dependent on which type of discrete chromatic dispersion compensation is implemented in the Rx. It is possible to mitigate the phase noise influence partly using a "radio frequency" (RF) pilot carrier [14-17] but the RF pilot carrier technique is not applicable for mitigating the EEPN caused phase noise [17].

Systems using pre-compensation of the dispersion (in the transmitter (Tx)) has been considered less frequently than post-compensation implementations, but initial demonstrations have been published [18,19]. The detailed influence of phase noise for pre-compensated implementations has – to the knowledge of these authors – not been investigated so far. In this paper, we compare the performance of QPSK systems using post- and pre-compensation of the fiber dispersion as far as the influence of laser phase noise is concerned.

The principle of pre- and post-compensation of the dispersion is presented in section 2 of this paper. Section 3 reports simulation results which benchmark the phase noise influence. Section 4 gives conclusive remarks.

---

\* **Corresponding author:** Gunnar Jacobsen, Acreo AB, Electrum 236, SE-16440 Kista, Sweden. E-mail: gunnar.jacobsen@acreo.se.

## 2. Theory

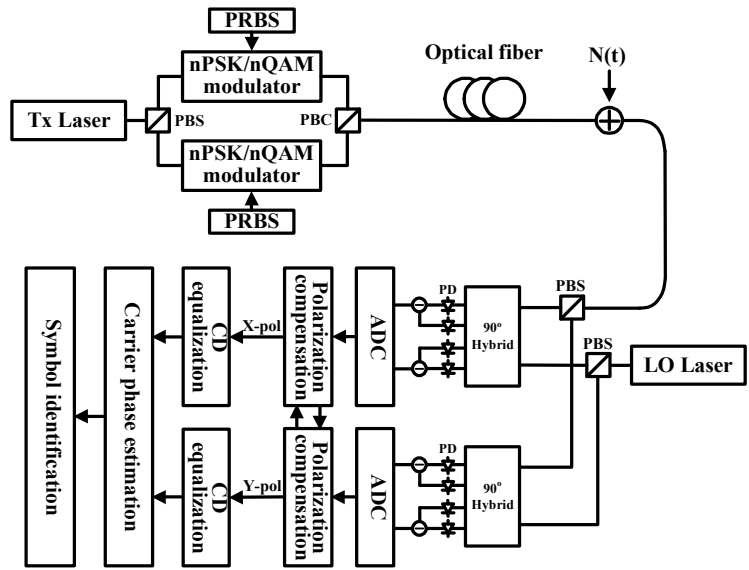

**Figure 1.** Block diagram for dual polarization nPSK/nQAM system using post-compensation of chromatic dispersion influence. *N(t)* shows the added optical noise which is used to measure the bit error rate (*BER*) as a function of optical signal-to-noise ratio (*OSNR*). Figure abbreviations: Tx – transmitter; PBS – polarizing beam splitter; PBC – polarization beam combiner; PRBS – pseudo random bit sequence; LO – local oscillator; ADC – analogue to digital conversion; CD – chromatic dispersion.

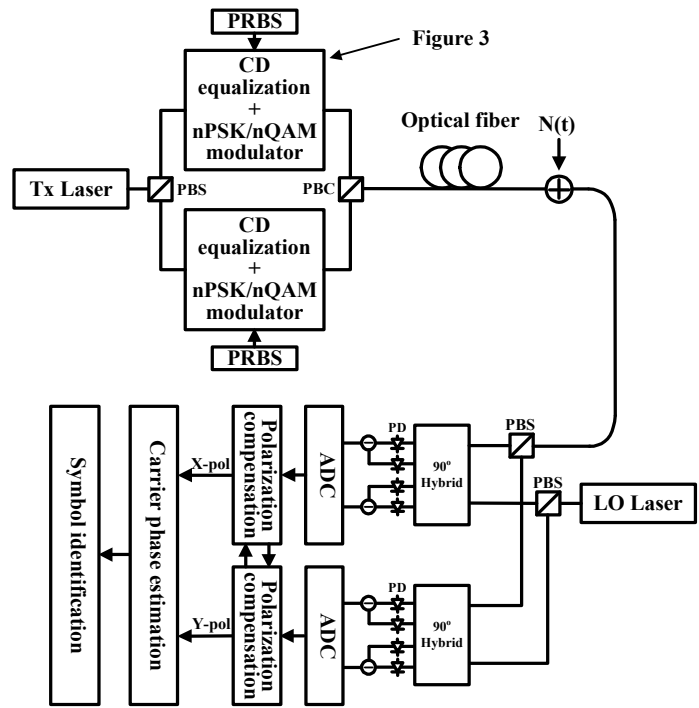

**Figure 2.** Block diagram for dual polarization nPSK/nQAM system using pre-compensation of chromatic dispersion influence. The transmitter implementation is shown in detail in Figure 3. *N(t)* shows the added optical noise which is used to measure the bit error rate (*BER*) as a function of optical signal-to-noise ratio (*OSNR*). Figure abbreviations: Tx – transmitter; PBS – polarizing beam splitter; PBC – polarization beam combiner; PRBS – pseudo random bit sequence ; LO – local oscillator; ADC – analogue to digital conversion; CD – chromatic dispersion.

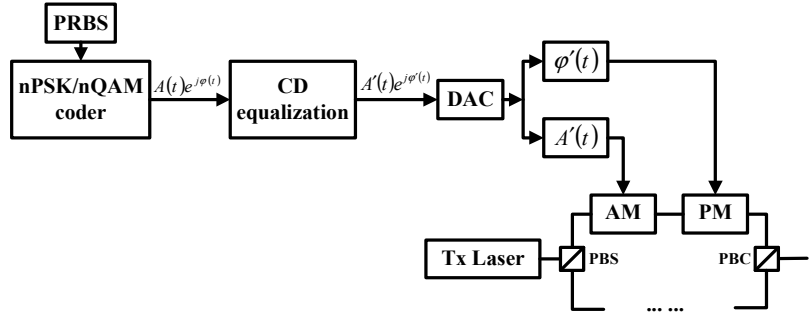

**Figure 3.** Transmitter for dual polarization nPSK/nQAM system using pre-compensation of chromatic dispersion influence. The nPSK/nQAM modulation is indicated in the time domain by $A(t)\cdot\exp(j\varphi(t))$ and the CD equalized signal by $A'(t)\cdot\exp(j\varphi'(t))$. A detailed discussion is given in section 2.2. Figure abbreviations: Tx – transmitter; PBS – polarization beam splitter; PBC – polarizing beam combiner; PRBS – pseudo random bit sequence ; DAC – digital to analogue conversion; CD – chromatic dispersion.

## 2.1 Post-compensation systems

It is adequate to discuss the total phase noise influence in the system – especially the equalization enhanced phase noise (EEPN) turns out to be important to account for in the design of high capacity, high constellation long range coherent optical systems. A system schematic for a dual-polarization coherent system employing post-compensation of the chromatic fiber dispersion is presented in Figure 1. The system schematic is for general type multilevel PSK or QAM modulation. When it comes to simulation examples in Section 3 we will focus on QPSK modulated systems.

The EEPN originates from dispersion interacting with laser phase noise. For the transmitter (Tx) laser in Figure 1 the net-dispersion originating from the transmission fiber and from the CD equalization in the Rx is zero and accordingly no EEPN results. The Local Oscillator (LO) laser is not influenced by the transmission fiber dispersion and the net-dispersion originates from the CD equalization. This results in EEPN from the LO laser [7]. The EEPN scales linearly with the accumulated chromatic dispersion and the linewidth of the LO laser. The variance of the additional noise due to the EEPN can be expressed as follows, see e.g. [7]:

$$\sigma^2_{EEPN} = \frac{\pi\lambda^2}{2c}\cdot\frac{D\cdot L\cdot \Delta f_{LO}}{T_S} \equiv 2\pi\Delta f_{EE}\cdot T_s \tag{1}$$

where $\lambda$ is the central wavelength of the transmitted optical carrier wave, $c$ is the light speed in vacuum, $D$ is the chromatic dispersion coefficient of the transmission fiber, $L$ is the transmission fiber length, $\Delta f_{LO}$ is the 3-dB linewidth of the LO laser, $\Delta f_{EE}$ is the 3 dB linewidth associated with EEPN and $T_S$ is the symbol period of the transmission system. This enables a definition of the effective intermediate frequency (IF) linewidth [13] - which defines the final phase noise influence in the receiver:

$$\Delta f_{Eff} \approx \frac{\sigma^2_{Tx}+\sigma^2_{LO}+\sigma^2_{EEPN}}{2\pi T_S} = \Delta f_{Tx}+\Delta f_{LO}+\Delta f_{EE} \tag{2}$$

where $\Delta f_{Tx}$ is the 3-dB transmitter laser linewidth, $\sigma^2_{Tx}=2\pi\Delta f_{Tx}\cdot T_S$ is the transmitter laser phase noise variance and $\sigma^2_{LO}=2\pi\Delta f_{LO}\cdot T_S$ is the LO laser phase noise variance. Eq. (7) implies that correlation between the LO and EEPN phase noise contributions can be neglected which is a valid approximation for a normal transmission fiber for very short (few km) or longer distances above the order of 80 km [21].

For the effective system linewidth given in (2) a Bit-Error-Rate (BER) floor results when the BER is plotted versus Optical-Signal-to-Noise-Ratio (OSNR). The BER floor level is given by [13]

$$BER_{floor} \approx \frac{1}{2}erfc\left(\frac{\pi}{4\sqrt{2}\sigma_{Eff}}\right) \tag{3}$$

where $\sigma^2_{Eff} \equiv 2\pi\Delta f_{Eff}\cdot T_S$ and *erfc* denotes the complementary error function.

## 2.2 Pre-compensation systems

A system schematic for a dual-polarization coherent system employing pre-compensation of the chromatic fiber dispersion is presented in Figure 2 and the detailed Tx implementation of the CD equalization is shown in Figure 3.

It is significance to comment on Figure 3 in some detail. The figure shows how the PSK/QAM modulation is generated in the electrical (time) domain. The general time specification of the analogue modulated signal (over one symbol period of duration $T_s$) is denoted $A(t)\cdot\exp(j\varphi(t))$ where $A(t)$ denotes the amplitude part of the modulation and $\varphi(t)$ denotes the phase part. In the case of PSK modulation we have $A(t)=1$. CD equalization (with a filter which has the inverse transfer function of the fiber dispersion [3,18,19]) followed by digital to analogue (DAC) conversion generates the signal $A'(t)\cdot\exp(j\varphi'(t))$ which is used to drive the amplitude modulator (AM) and phase modulator (PM). This moves the PSK/QAM modulated signal onto the optical carrier wave. Compared to the post-compensation case (Figure 1) it is observed that the CD equalization in the transmitter is performed in the electrical domain prior to the optical signal generation. This means that the dispersion of the CD equalizing filter does not interact with the Tx laser phase noise. Because of this the net-dispersion for the EEPN generating parts of the system is the fiber for the Tx laser whereas there is no dispersive system parts which interacts with the LO laser. This means that the resulting EEPN originates from the Tx laser and is given by (1) when $\Delta f_{LO}$ is replaced by $\Delta f_{Tx}$.

## 3. Simulation results and discussion

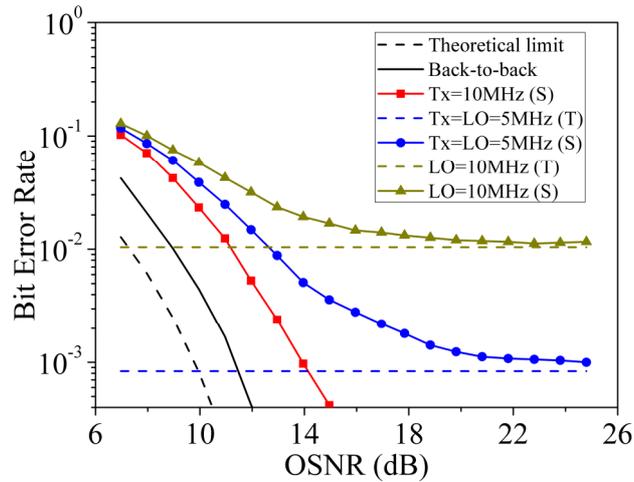

**Figure 4.** BER for single polarization QPSK coherent system using post-compensation of chromatic dispersion. Results are for a transmission distance of 2000 km. Different cases are indicated in the figure. Tx and LO laser linewidths are indicated in the figure. The PRBS length is $2^{16}-1$. Figure abbreviations: OSNR - optical signal-to-noise ratio; Rx – receiver; RF – radio frequency; S – simulation results; T – BER floor using (3).

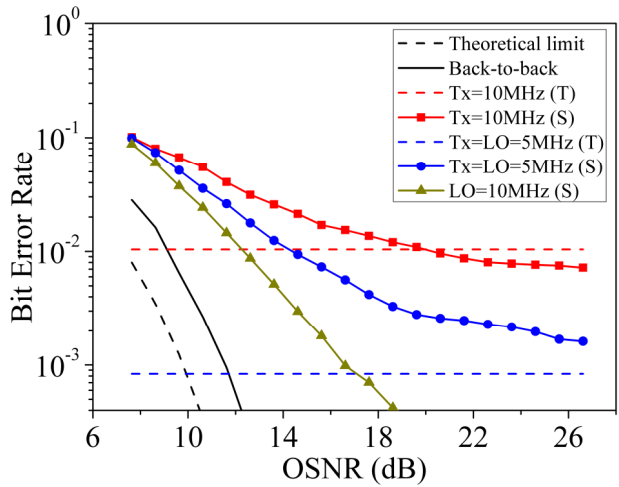

**Figure 5.** BER for single polarization QPSK coherent system using pre-compensation of chromatic dispersion. Results are for a transmission distance of 2000 km. Different cases are indicated in the figure. Tx and LO laser linewidths are indicated in the figure. The PRBS length is $2^{16}-1$. Figure abbreviations: OSNR - optical signal-to-noise ratio; Rx – receiver; RF – radio frequency; S – simulation results; T – BER floor using (3).

In this section we will give simulation results for a 56 Gb/s single polarization QPSK coherent system i.e. the symbol rate is 28 GS/s which is modeled using one polarization branch of the system indicated in Figure 1. We consider a system where the CD equalization is performed by finite impulse response (FIR) filtering with a number of active tabs which are adjusted according to the amount of chromatic dispersion (the transmitted fiber length) – see [3]. The transmission fiber is a normal single-mode fiber with dispersion coefficient of *D=16* psec/nm/km. We have 2 signal samples per symbol period. A single tab least mean square (LMS) filter is used in the Rx for carrier phase estimation and noise filtering [20]. The Tx and LO laser phase noise is modeled assuming a white frequency noise power spectral density i.e. the phase noise has Gaussian probability density function (see e.g. [21]). The simulation is performed within the VirtualPhotinics system simulation software environment [22].

Figure 4 shows the Bit-Error-Rate (BER) versus Optical Signal-to-Noise-Ratio (OSNR) for 2000 km transmission distance for the post-compensation system. This is a situation where the EEPN is dominating. For a value of the LO laser linewidth $\Delta f_{LO}$ of 5 MHz we have an EEPN linewidth $\Delta f_{EE}$ of 206 MHz. For $\Delta f_{LO}=10$ MHz the EEPN linewidth is 412 MHz. Figure 4 shows BER results for a total laser linewidth ($\Delta f_{Tx}+\Delta f_{LO}$) of 10 MHz and for 3 situations where 1) $\Delta f_{Tx}=10$ MHz, $\Delta f_{LO}=0$; 2) $\Delta f_{Tx}=\Delta f_{LO}=5$ MHz and 3) $\Delta f_{LO}=10$ MHz, $\Delta f_{Tx}=0$. From the figure it appears clearly that when the LO laser linewidth is zero (and we have no EEPN influence) the BER-floor is well below $10^{-3}$. When we have the largest EEPN influence (for the LO-linewidth of 10 MHz) the BER-floor is around $10^{-2}$ and we have excellent agreement with the prediction of (3). For the linewidth of 5 MHz for both lasers we have a BER floor of around $10^{-3}$ and again excellent agreement between simulations and (3).

Figure 5 shows the BER results versus OSNR for the pre-compensation system implementation with the same 3 cases as for the post-compensation system considered in Figure 4. When comparing Figure 4 and 5 it is obvious that they are very similar except for the fact that in the pre-distortion implementation the EEPN results from the Tx laser linewidth. We also observe a slightly poorer agreement between the BER-floor predicted by (3) and the simulation results for the pre-compensation case. This is tentatively attributed to the fact that in the post-compensation case the EEPN is generated by the discrete CD equalization filter in the Rx (and this is the basis for the specification of the resulting phase noise variance in (1) which is derived in [7]) whereas in the pre-compensation case the EEPN is generated in the analogue domain (by the transmission fiber dispersion) which is a somewhat different physical phenomenon.

It is obvious that the electrical CD equalization is the origin of the EEPN influence for both pre- and post-compensation implementations. This is a major difficulty in the practical implementation of longer-range optical coherent high capacity high constellation systems. It is worth to note that this difficulty can be avoided by using pure optical CD equalization i.e. using dispersion compensating fibers (CDFs). The use of CDFs will completely eliminate the EEPN generation and seems to be an obvious choice for the practical implementation of advanced optical coherent longer-range systems.

## 4. Conclusions

We have presented a comparative study in order to specify the influence of equalization enhanced phase noise (EEPN) for pre- and post-compensation of d chromatic dispersion in high capacity and high constellation systems. This is – to our knowledge – the first detailed study in this area for pre-compensation systems. Our main results show that the local oscillator phase noise determines the EEPN influence in post-compensation implementations whereas the transmitter laser determines the EEPN in pre-compensation implementations. As a result of significance for the implementation of practical longer-range systems it is to be emphasized that the use of chromatic dispersion equalization in the optical domain – e.g. by the use of dispersion compensation fibers – eliminates the EEPN entirely. Thus, this seems a good option for such systems operating at high constellations in the future.